# Personalized Real Time WeatherForecasting With Recommendations
A new era of weather forecasting

Abhishek Kumar Singh
Assistant Professor, CSE, BIT
Durg, C.G., India

Aditi Sharma
Assistant Professor,CSE, JIIT
Noida, India

Rahul Mishra
P.Hd. Scholar
IIIT-Delhi

*Abstract*—Temperature forecasting and rain forecasting in today's environment is playing a major role in many fields like transportation, tour planning and agriculture. The purpose of this paper is to provide a real time forecasting to the user according to their current position and requirement.

The simplest method of forecasting the weather, persistence, relies upon today's conditions to forecast the conditions tomorrow i.e. analyzing historical data for predicting future weather conditions. The weather data used for the DM research include daily temperature, daily pressure and monthly rainfall.

## I. INTRODUCTION

Forecasting the temperature and rain on a particular day and date is the main aim of this paper. In the paper we forecast rain and temperature for Europe; year up to 2051 and also we forecast temperature of world; year up to 2100.Our paper is aimed to provide real time weather forecast service at finest granularity level with recommendations. We grab user's location (longitude, latitude) using GPS data service whenever user requests for our services. Our system will process the users query and will mine the data from our repository to draw appropriate results. Users will be provided with recommendations also and that is the key facility of our service. Personalized forecast is generated for each individual user based on their location.

## II. SCOPES

The project mainly focuses on forecasting weather conditions using historical data. This can be done by extracting knowledge from this given data by using techniques such as association, pattern recognition, nearest neighbor etc.

- Disaster Mitigation: Predicting storms, floods, droughts
- Helping those sectors which are most dependent on weather such as agriculture, aviation also depends on weather conditions.

## III. TARGET SEGMENT

Some target segments are following.

*1) Our target users are mainly normal citizens they can use our services for their lots of benefits like:*

*a) Suppose a user is stuck on the way to home due to heavy rain then using our service, he will be able to know whether there is any another highway or route nearby where it's not raining or less raining.*

*2) Using our service any individual can get weather information specially personalized for him irrespective of what is the time or place.*

*3) If our service is connected with THERMOSTATs of some house then temperature of the house can be controlled automatically using forecast information provided by us using location of house based on GPS.*

*4) In transportation industry our services can be used to take some important decisions like: Which route is better for transportation, where snow fall probability is quite low etc.*

*5) There are enormous more areas where our service will be helpful like tourism, food processing industry, Aviation Industry, Oil and natural gas exploration and production activities etc.*

## IV. PROCESS DETAILS:

The process we have briefed in earlier section can be depicted pictorially and which is self-explanatory.

We can divide our process in two modules namely:
*1) Weather Mining*
*2) Recommendation*

*B. Weather Mining:*

- Data collection: We have collected weather data from WORLD DATA CENTER for climate, Hamburg. We have decided to use NWS API for data collection in future.

- Data formatting and cleaning: We have converted our data from .NC (netcdf) format to .CSV (comma-separated values) format because WEKA supports .CSV format.

- Clustering: Using WEKA, we have performed clustering on weather data to draw inferences.

- Recommendation: We have planned to use recommendation algorithm as user to locationcollaborative algorithm similar to user to item collaborative algorithm. This algorithm uses user location (N*M) metrics.

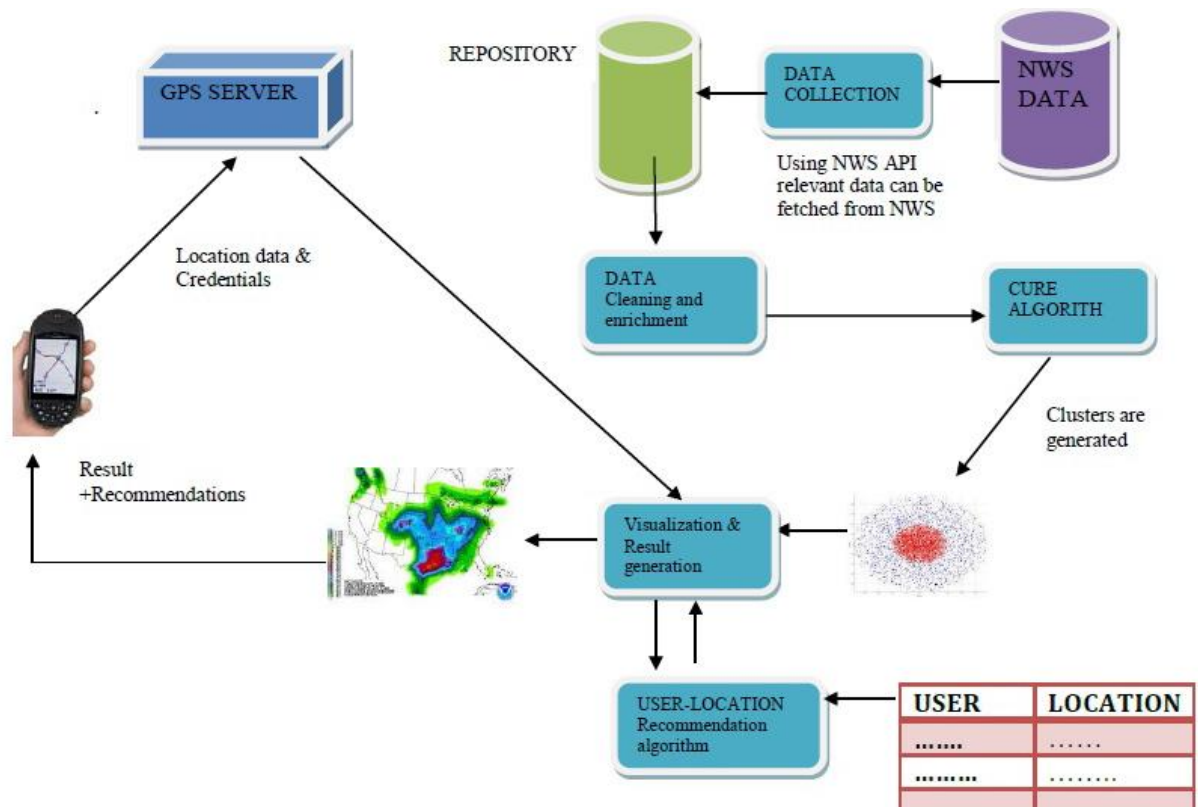

Fig. 1. Process details

- Visualization: To generate visualization for user, we have used NOAA weather and climate tool kit.

For the part of the implementation, on which your project focused most, which algorithms you implemented or used and if any modifications were needed to those algorithms or if you did some initial preprocessing, discuss here For the other phases of data mining, discuss briey. E.g., if you focused most on visualization, you can talk about: which data (Example: downloaded from some website put the URL here; did some survey, then talk about how you did the survey etc) collection approach was used in the project?

*C. Recommendation*

- Extract the location of the user
- Extract the destination of the user
- And then recommend the best path according to the conditions. (As shown in figure 1)

## V. TECHNICAL SKILLS

We haven't implemented software for clustering and visualization our self, we have used WEKA and NOAA for this purpose. We have faced some problems to work with these tools and eventually solved them on our own.

## VI. SOME SNAPSHOTS:

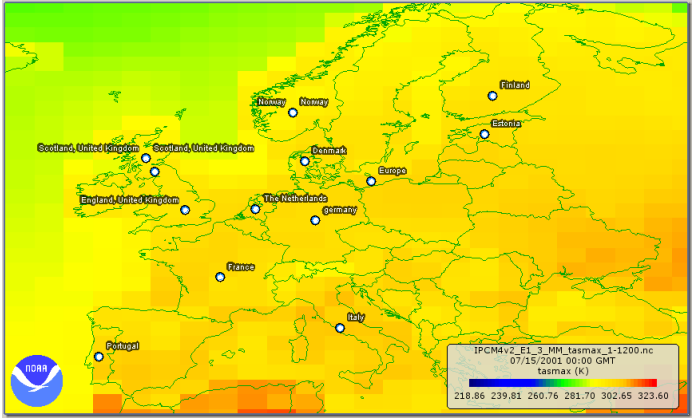

Fig. 2. Temperature Forecast for Europe in 7/2001

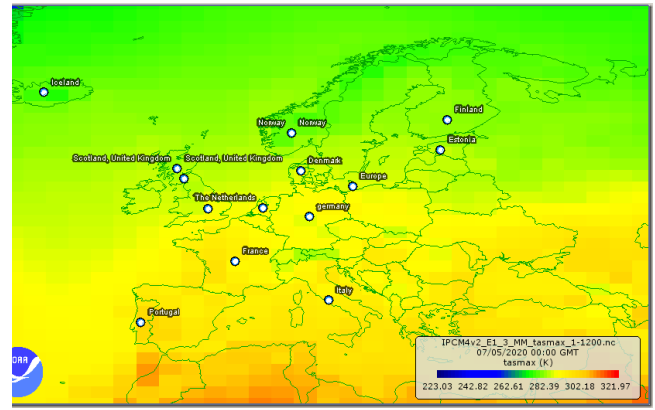

Fig. 3. Temperature Forecast for Europe in 7/2020

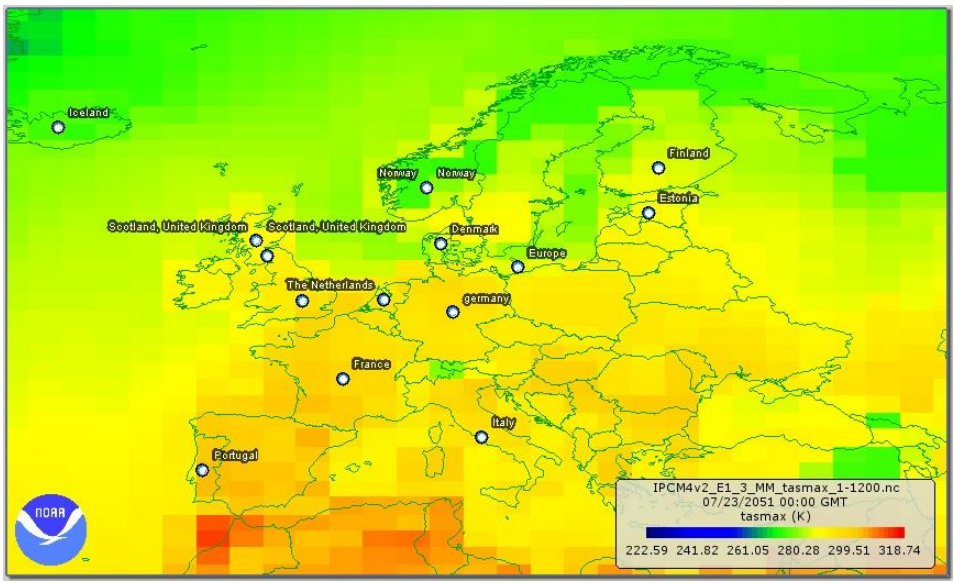

Fig. 4. Temperature Forecast for Europe in 7/2051

- WEKA was crashing initially for large data set.
  - ➢ We resolved this issue by increasing heap size of WEKA.
- Since WEKA doesn't support netcdf format. We had to convert our data to .CSV format.
  - ➢ We have written code in C++ to convert data from .NC to .CSV.

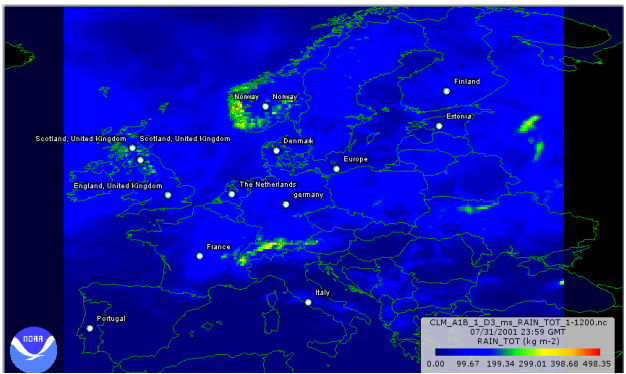

Fig. 5. Rain Forecast for Europe in 7/2001

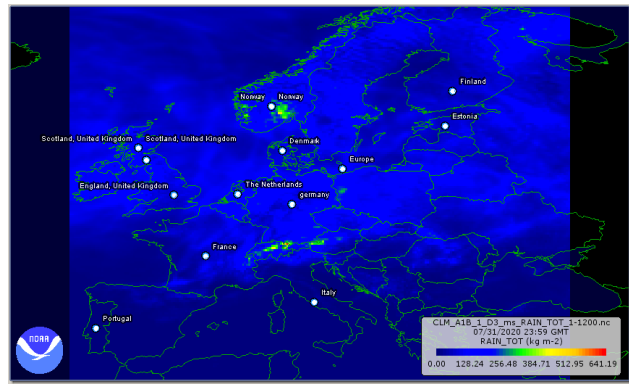

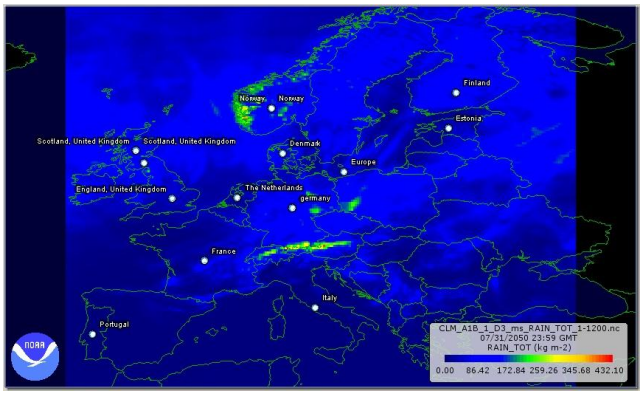

Fig. 7. Rain Forecast for Europe in 7/2050

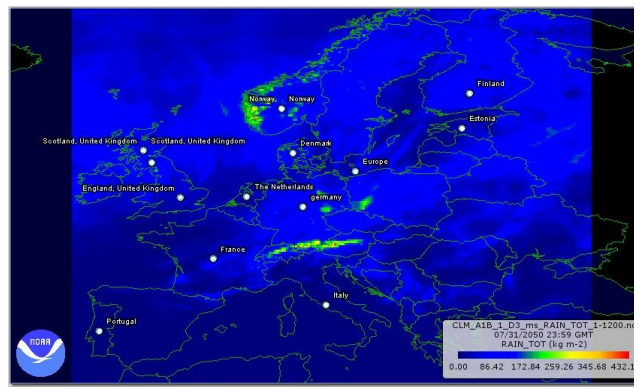

Fig. 8. Rain Forecast for Europe in 7/2100

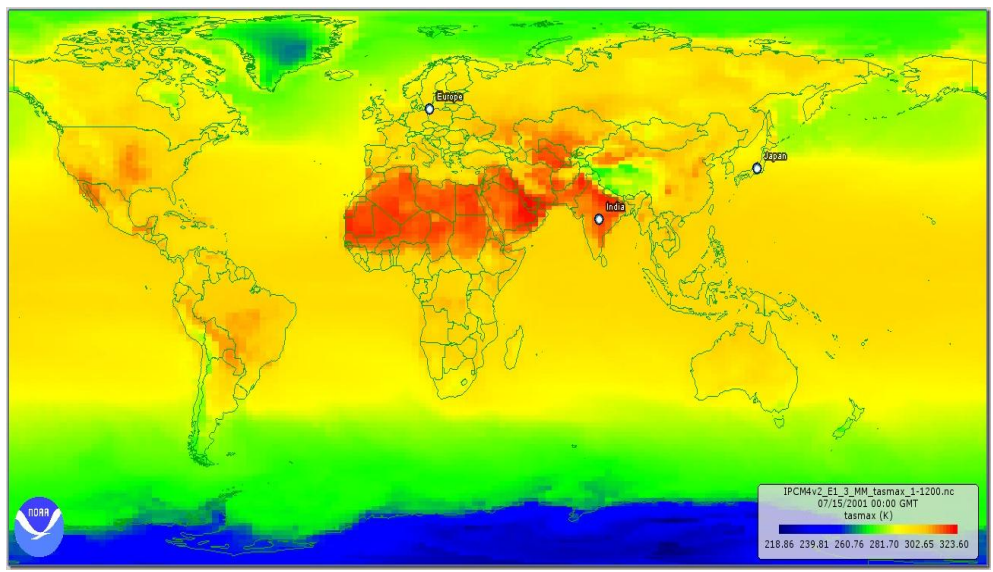

Fig. 9. Temperature Forecast for world in 7/2001

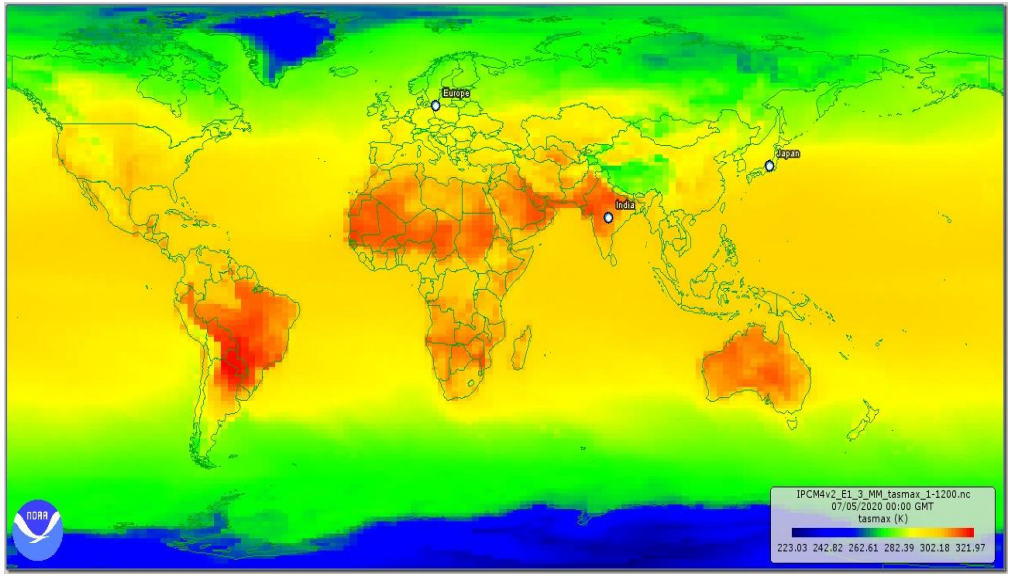

Fig. 10. Temperature Forecast for world in 7/2020

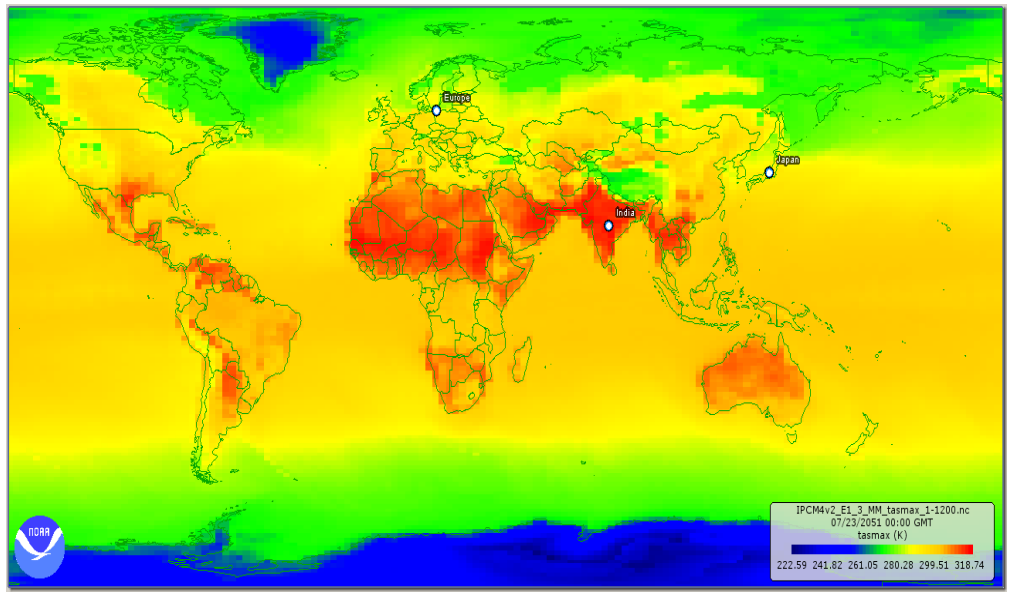

Fig. 11. Temperature Forecast for world in 7/2061

REFERENCES

[1] Weka tool.
[2] NOAA tool.
[3] World Climate Data Center for Data